\documentclass[runningheads]{llncs}
\usepackage{url}
\usepackage[pdftex]{graphicx}
\usepackage{todonotes}
\usepackage{float}
\usepackage{booktabs}
\usepackage[pdftex,bookmarks=true,plainpages=false,pdfpagelabels=true]{hyperref}
\usepackage{enumerate}
\usepackage{paralist}
\usepackage{array}
\usepackage{longtable}
\usepackage{listings}
\usepackage[utf8]{inputenc}
\usepackage[capitalize, noabbrev]{cleveref}
\usepackage{tabularx}
\usepackage{multirow}
\usepackage{bigstrut}
\usepackage{verbatim}
\usepackage{geometry}

\newcolumntype{L}[1]{>{\raggedright\let\newline\\\arraybackslash\hspace{0pt}}p{#1}}
\newcolumntype{C}[1]{>{\centering\let\newline\\\arraybackslash\hspace{0pt}}p{#1}}
\newcolumntype{R}[1]{>{\raggedleft\let\newline\\\arraybackslash\hspace{0pt}}p{#1}}
\newcommand{\quotes}[1]{``#1''}

\newcommand*{\nssafe}{{S\textsuperscript{2}C-SAFe}}
\newcommand*{\ssafe}{{S\textsuperscript{2}C-SAFe}~}

\begin{document}
\title{How to Integrate Security Compliance Requirements with Agile Software Engineering at Scale? Authors' Copy}
\titlerunning{Authors' Copy}
\author{Fabiola Moy\'on\inst{1} \orcidID{0000-0003-0535-1371}
\and  
Daniel M\'endez Fern\'andez \inst{2} \orcidID{0000-0003-0619-6027}  
\and \\  
Kristian Beckers\inst{3} 
\and   
Sebastian Klepper\inst{4} } 
\authorrunning{F. Moy\'on et al.}
\authorrunning{Authors' Copy}
\institute{Technical University of Munich and Siemens, Germany \email{fabiola.moyon@tum.com}
\and
Blekinge Institute of Technology, Sweden, and fortiss GmbH, Germany \email{daniel.mendez@bth.se} 
\and
Social Engineering Academy, Germany \email{kristian.beckers@social-engineering.academy}
\and
Technical University of Munich, Germany \email{sebastian.klepper@tum.de}
}
\maketitle              % typeset the header of the contribution
\begin{abstract}
Integrating security into agile software development is an open issue for research and practice. Especially in strongly regulated industries, complexity increases not only when scaling agile practices but also when aiming for compliance with security standards.
To achieve security compliance in a large-scale agile context, we developed \nssafe: An extension of the Scaled Agile Framework that is compliant to the security standard IEC~62443-4-1 for secure product development.

In this paper, we present the framework and its evaluation by agile and security experts within Siemens' large-scale project ecosystem. We discuss benefits and limitations as well as challenges from a practitioners' perspective. Our results indicate that \ssafe  contributes to successfully integrating security compliance with lean and agile development in regulated environments. We also hope to raise awareness for the importance and challenges of integrating security in the scope of Continuous Software Engineering.

\end{abstract}
\section{Introduction}\label{sec:intro}
Security compliance is a major concern for several industries~\cite{Bell:2017,Fitzgerald:2013}. Typically, security practitioners (and regulators) hold a holistic view on security affecting people, processes, and technology~\cite{IEC11,Bell:2017,HumpreysISO}. The perspective of practitioners, however, is rather dispersed and security is commonly treated as just another non-functional requirement~\cite{Fitzgerald:2017}. Security engineering activities are further too often applied in an ad-hoc manner to a limited set of security problems, e.g., vulnerability testing or static code analysis~\cite{Bell:2017}. Security concerns are often mixed with software functionality and limited to specific implementations like authentication or encryption~\cite{Jaatun:2017}.

Integrating security into lean and agile processes further intensifies these issues and constitutes a well-known research problem~\cite{Turpe:2017,Fitzgerald:2017,ahola_handbook_2014}. This is especially true for large software development projects. One challenge here is to fulfil requirements rigorously to comply with regulations while not limiting the speed and flexibility agile development methodologies promise. However security standards often require a series of processes to define, analyse, and mitigate security vulnerabilities~\cite{ISO27001} whereas lean and agile methodologies aim at avoiding rigid linear processes. While the agile manifesto states \quotes{to value individuals and interactions over processes}, \quotes{collaboration over contract negotiation}, and \quotes{responding to change over following a plan}~\cite{Beck:2001}, standards explicitly demand documented evidence of responsibilities, agreements, and established development procedures.

Our research shall provide a perspective for resolving this conflict through \emph{Continuous Security Compliance}. In particular, we aim at implementing security standard requirements along with agile development methodologies. To this end, we analysed the issue in a large industrial setting and its currently applied norms: the Scaled Agile Framework (SAFe) as well as the IEC~62443-4-1 standard, later we propose a revised framework dubbed \ssafe. We chose the IEC~62443-4-1 standard for secure product development, released in 2018 based on previous secure product development standards such as BSIMM \cite{BSIMM}, ISO27034\cite{ISO27034}, or Security by Design with CMMI~\cite{SCMMI}. 
Our framework shall maintain SAFe's perspective on development procedures and principles while capturing the essential requirements of security standards. In this paper, we contribute:
\begin{enumerate}
	\itemsep=-1pt
	\item The proposal of our \ssafe framework, a security-standard compliant variant of the Scaled Agile Framework. %[\textbf{RQ~1}] Is there a viable approach to integrate security standards into agile development?
%[\textbf{RQ~2}] Can the scaled agile framework comply with the IEC 62443-4-1 standard? 
	\item An evaluation of the \ssafe framework in large-scale software development environments. Given that the introduction of SAFe may take up to 8 years in the chosen organisational context, we conduct our evaluation in a preliminary manner focusing particularly on expert interviews.

%	\item [\textbf{RQ~3}] Which are the challenges for achieving security compliance in scaled agile projects? 
\end{enumerate}

We conclude our evaluation with the practitioners' perception of the challenges to achieve security compliance in a continuous manner. By sharing these insights, we particularly hope to raise awareness for the importance, but also challenges of integrating security in large-scale software development organisations following lean and agile principles.

\section{Fundamentals and Related Work}
\label{sec:relatedWork}
% Continuous Software Engineering

\emph{Continuous Software Engineering } (CSE) utilises lean and agile principles for a rapid and continuous ``flow'' of activities across business, development, and operations~\cite{Fitzgerald:2015}. 

In their ``Continuous *'' model of CSE, Fitzgerald et al.~\cite{Fitzgerald:2017} describe Continuous Security and Continuous Compliance as related but separate concerns and activities.
\emph{Continuous Compliance} (CC) seeks to satisfy regulatory compliance standards on a continuous basis rather than a ``big-bang'' approach to ensure compliance at release time~\cite{Fitzgerald:2013,McHugh:2013}. \emph{Continuous Security} (CS) elevates security from non-functional requirement to key concern by efficiently identifying and addressing security issues throughout all processes~\cite{Fitzgerald:2015}.

Related work discusses the suitability of agile methods for regulated environments~\cite{Fitzgerald:2013} or the extensibility of their use~\cite{Cawley:2010}. With regard to security, authors focus on solving security aspects in agile environments, without considering regulations as focus \cite{Beznosov:2004,Siponen:2005,Bartsch:2011,Baca:2012}; or deriving security activities from a regulations perspective but lacking attention to lean and agile environments as well as corporate operating procedures, e.g., product life cycle~\cite{Cawley:2010,Beckers:2015}. 
Practical concerns of CS are: adapting the development process to security, better eliciting and tracking security requirements, and incorporating assurance into iterations~\cite{Bartsch:2011}. 

Separating CS and CC is illustrated by Fitzgerald et al.~\cite{Fitzgerald:2013}, concluding that agile methods are suitable for security-critical environments, but not yet adopted in regulated environments .

% Continuous Security Compliance

We aim for \textit{Continuous Security Compliance} (CSC): combining CC and CS through the holistic view of standardisation that spans across people, processes, and technology~\cite{IEC11}. Regulatory requirements are utilised to derive security activities and therefore integrating security into a process while also making it standards-compliant~\cite{Moyon:2018}. Further work concentrates on security governance best practices~\cite{Daennart:2019}.  
This is complementary to prior work focused on the technology side, integrating security engineering into agile processes~\cite{Felderer:2017,Bell:2017,ahola_handbook_2014,Baca:2011,Choliz:2015}, or on the process side, integrating desirable but not standards-compliant security activities~\cite{ahola_handbook_2014,Baca:2015,Stephanow:2017}.

% Contribution

\ssafe is the result of applying this holistic principle to both a security-critical and a regulated domain: industrial and automation control systems. The result is an in-depth analysis of a security standard (IEC~62443-4-1) followed by the integration with lean and agile development practices represented by the Scaled Agile Framework (SAFe).

\emph{IEC~62443} constitutes a series of standards for network and system security published by the International Electrotechnical Commission (IEC). The standard focuses on requirements for component providers for industrial automation and control systems (IACS), part 4-1 describes process requirements for secure product development~\cite{IEC:2017}. We reference this part of the standard as \quotes{4-1} or \quotes{4-1 standard}.
\emph{SAFe} is a widely used process framework that scales lean and agile development to large organisations with multiple levels. It furthermore defines the corresponding roles, responsibilities, activities, and artefacts~\cite{Leffingwell:2017}.

For such IACS environments our contribution aims to bridge the gap between lean and agile development, practical security, and compliance~\cite{Jaatun:2017}.

\section{\texorpdfstring{\ssafe Framework in a Nutshell}{S2C-SAFe Framework in a Nutshell}}
\label{sec:method}
\begin{figure}[ht]
	\centering
	\includegraphics[width=0.4\textwidth]{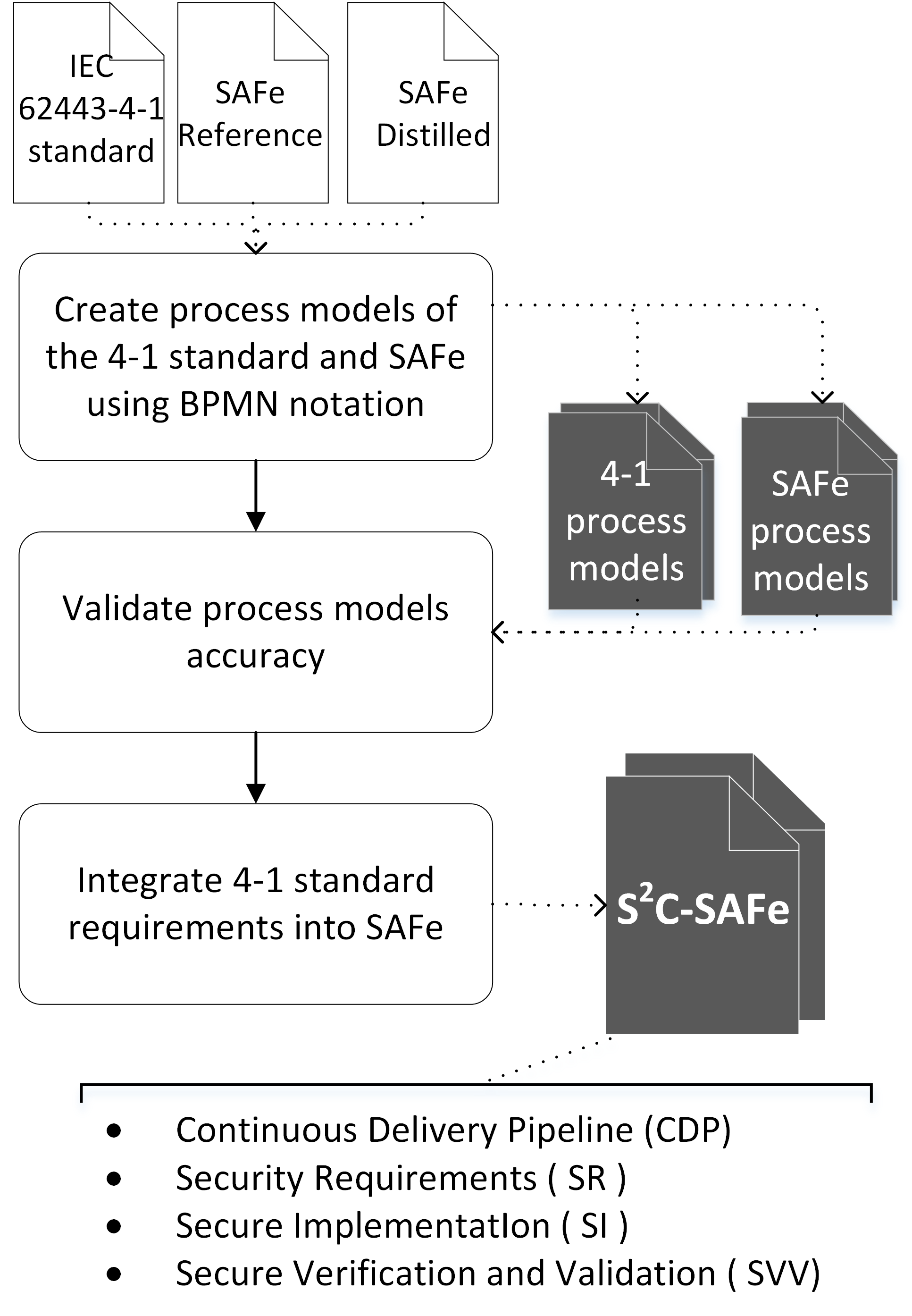}
	\small \caption[Creation of \ssafe]{Creation of \ssafe by generating and merging visual models of 4-1 and SAFe. Black document symbols designate our contribution. In previous work, we described the integration method~\cite{Moyon:2018}. The present contribution presents the \ssafe framework and its evaluation.}
	\label{fig:method}
\end{figure}

The overall aim of our work is to improve product development life-cycle by integrating requirements of IEC~62443-4-1 into SAFe, resulting in the ``Security Standard Compliant Scaled Agile Framework'' (\ssafe). \Cref{fig:method} shows how this is achieved by using visual modelling and by merging techniques as presented in our previous work~\cite{Moyon:2018}.
Essential elements of SAFe and 4-1, such as roles, activities, and artefacts, were captured using Business Process Model and Notation (BPMN), a visual modelling language capable of expressing all of these aspects at once. After refining these models separately with expert practitioners, the process framework model is extended with elements from the security standard model, yielding the \ssafe framework. Previously we found that a visual approach allows for more focused reviews than textual representation.

\begin{figure}[!ht]
	\centering
	\includegraphics[width=0.9\textwidth]{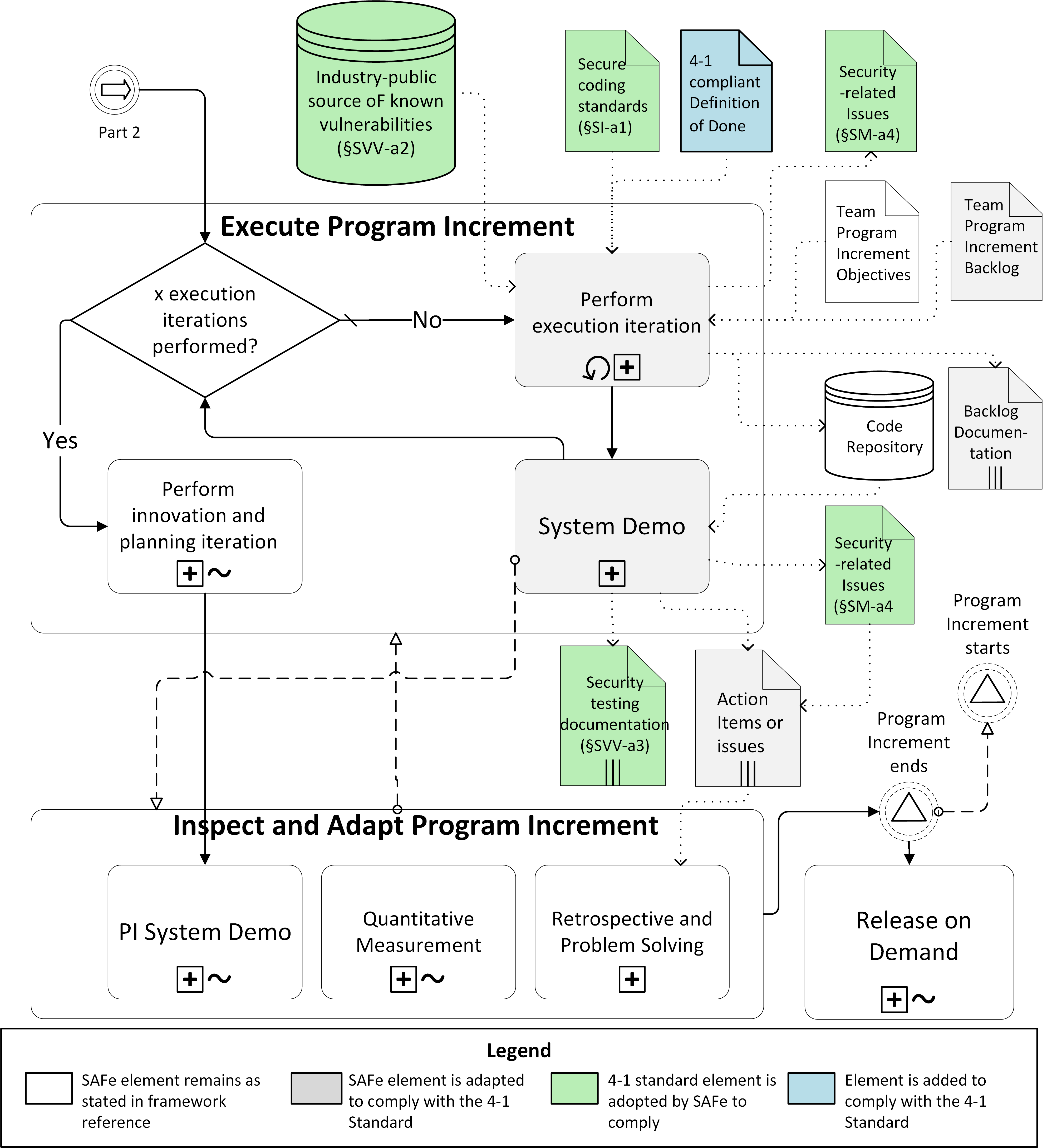}
	\small \caption{Excerpt of \ssafe Continuous Delivery Pipeline (CDP). This overview model describes the processes involved to execute and inspect a program increment as described in SAFe plus the artefacts required by the 4-1 standard in the practices of SR, SI, and SVV.}
	\label{fig:sol_CDP02}
\end{figure}

\ssafe describes how requirements of 4-1 can be implemented within SAFe by showing when to involve roles, execute activities, or generate artefacts. It focuses on SAFe's Continuous Delivery Pipeline (CDP), where the actual product development occurs, and makes it compliant with security requirements (SR), secure implementation (SI), and security verification and validation testing (SVV). These scopes address concerns we captured from practitioners such as frequent vulnerability testing, security requirements traceability, or coding standards review. In addition to a CDP model integrated with SR, SI, and SVV, \ssafe contains detailed models for each practice.
\cref{fig:sol_CDP02} shows an overview of the \ssafe CDP. The full framework is available in the online material associated with this paper \footnote{\url{https://dx.doi.org/10.6084/m9.figshare.7149179}}.

\subsection{Security Requirements (SR)}

SAFe does not specify where and how to elicit security requirements even though (security) requirements elicitation constitutes a major challenge both in practice and research~\cite{Fernandez:2013:NPR}, especially when developing a product threat model and deriving requirements to counter threats~\cite{Bartsch:2011,Fernandez:2016}.
\ssafe therefore explicitly considers security requirements at program and team level and makes them part of the Backlog, equal to all other requirements in prioritisation and traceability. Security Experts facilitate analysis but are not primarily responsible. Instead, Product Management, Business Owners, and Systems Architects are in charge so they become aware of threats. Similarly, the Product Owner requires adequate training to be able to prioritise and approve security requirements.

%S2C-SAFe explains that Security Experts role is to facilitate security analysis but not to assume it as own function. When SAFe roles are in charge of security, they will understand threats. Later, prioritization of security requirements will follow same rules that other requirements and not be postponed. This problem was referred during initial discussions with practitioners (see Appendix C Figure C.1 and Figure C.2).
% Reading carefully S2C-SAFe will immediately alert over the need to train the Product Owner on security. He is in charge of prioritizing and approve them after implementation.

\subsection{Secure Implementation (SI)}

SI involves following secure coding standards to avoid vulnerabilities.
\ssafe follows a process based on coding analysis as introduced in~\cite{Baca:2011,Baca:2012,Baca:2015}. It defines coding standards early at program level during the PI Planning Event. Security Experts provide guidance so they suit domain and solution.
To ensure that coding standards are followed, they are made part of the Definition of Done and agile teams as well as the product owner are trained accordingly.

\subsection{Security Verification and Validation Testing (SVV)}

\begin{figure}[!ht]
	\centering
	\includegraphics[width=0.9\textwidth]{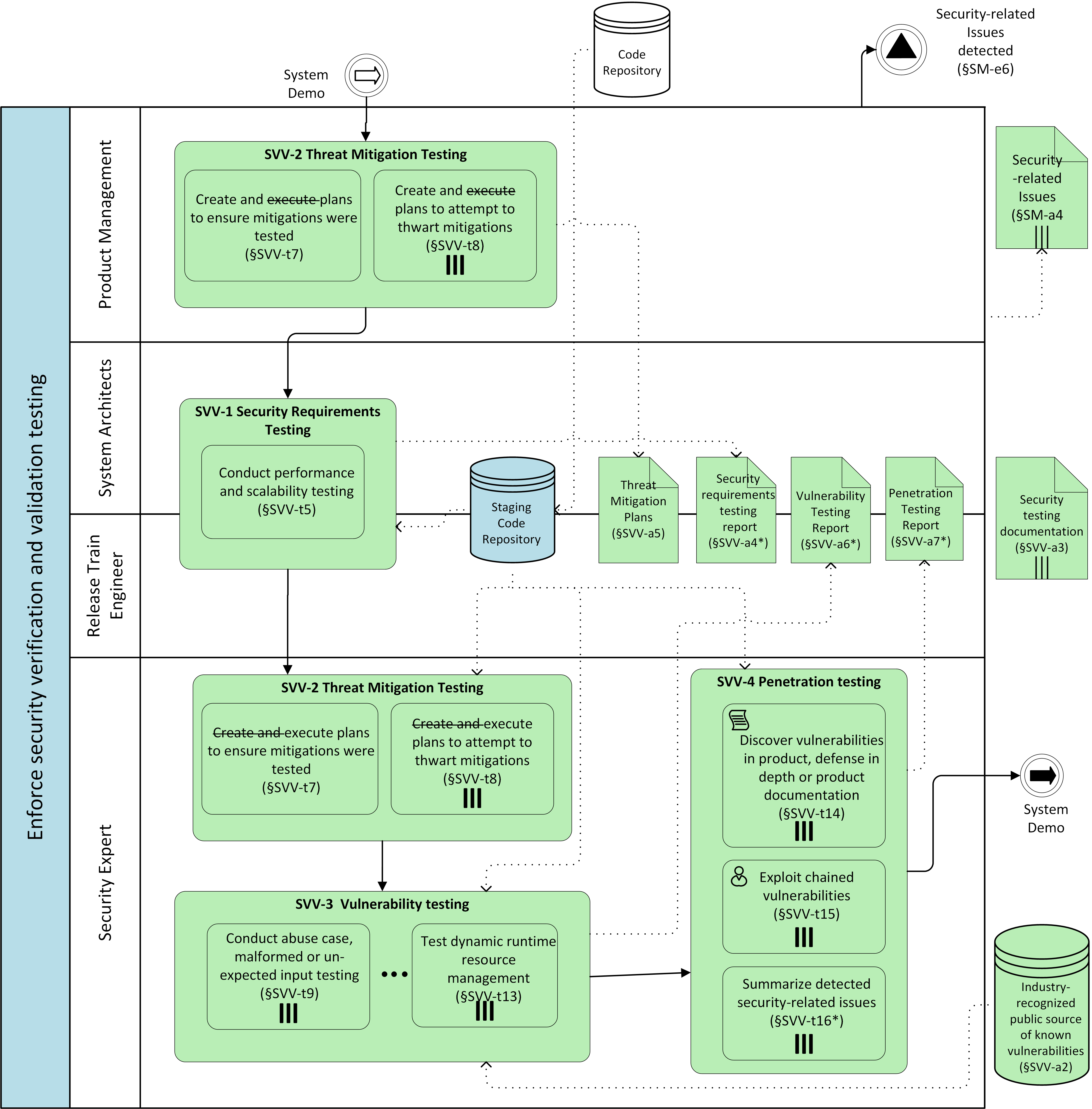}
	\small \caption{\ssafe System Demo refinement model. Process diagram that depicts a new activity for SAFe System Demo to perform security verification and validation testing. A \texttt{Security expert} participates for certain types of testing while SAFe Program level actors are also responsible of security testing. Color coding is consistent with \cref{fig:sol_CDP02}.}
	\label{fig:sol_svv_03}
\end{figure}

SVV focuses on detecting and resolving vulnerabilities.
One major concern is independence of testers which is enforced through independence rules during formation of agile teams.
\ssafe also defines how further activities such as security functionality testing, vulnerability testing, or penetration testing apply to features, user stories, or both.
It also defines criteria to keep resource allocation efficient and ensure continuous security testing, placing security functionality testing at team level and conducting all testing activities on program level before every System Demo. \ssafe contains models that shows a 4-1 compliant SAFe System Demo (see System Demo box in \cref{fig:sol_CDP02}). \Cref{fig:sol_svv_03} is a more granular refinement showing testing tasks and artefacts, as referred by the 4-1 practice SVV, and their mapping to SAFe roles. Further models are available in the online material.

\section{Study Design}
We evaluated \ssafe via expert interviews involving 16 practitioners working at Siemens in security compliance or (agile) software engineering. Among these experts are IEC committee members for 4-1 as well as SAFe core contributors.

Our overall goal is to explore the meaningfulness of our approach to the needs in a practical context characterised by security-critical and large-scale agile development of software or software-intensive systems. Our evaluation is guided by the following two research questions:

\begin{itemize}
	\item[\textbf{RQ~1}] From the perspective of practitioners, how applicable is \ssafe in this type of environment?
	\item[\textbf{RQ~2}] Which challenges do practitioners see when pursuing security compliance in this type of environment?
\end{itemize}

Our intention is to explore potential benefits and limitations of the here proposed framework. This shall lay the foundation for a roll-out that is minimally disruptive to the organisation and maximally intuitive for practitioners.

\subsection{Subject Selection}
In the industrial environment, where \ssafe is meant to be applied, projects are characterised by large-scale agile practices involving security experts on demand. Since industrial systems are part of critical infrastructure, such projects must comply with formal security standards, like the 4-1 standard when referring to product development. Such projects involve various agile teams with six people each. In those settings projects require direct cooperation between security experts and development teams.

%Typical project settings in which \ssafe is meant to be applied are characterised by large-scale agile practices involving security experts. As and the need for formal security compliance. Most projects that fit these characteristics involve various agile teams with six people each, e.g., in the industrial automation sector. In those settings projects require direct cooperation between security experts and development teams.

We consciously selected from both groups: development teams working in these settings and security experts joining those projects on-demand, e.g., in conjunction with internal audits.

As these are all experienced professionals, we defined profiles to distinguish their level of expertise according to their key role. \Cref{tab:6_evaluatorsPerProfileGroup} shows each role's background and share of our 16 interviews.
We distinguish top experienced subjects who contribute to the 4-1 standard (\emph{Contributor IEC}) or to the SAFe framework and its dissemination within the company (\emph{Contributor SAFe}). We further distinguish \emph{Principle Experts}, having vast knowledge and leading teams, \emph{Senior Experts}, having deep knowledge and guiding colleagues, and finally \emph{Experts} who are responsible for setting up specific topics into practice.

\begin{table}[!htbp]
\centering
	\caption[Mapping of interviews to profile groups]{Mapping of interviews to subject profile and background.}
	\scalebox{0.95}{
		\begin{tabular}{L{3cm}C{1.5cm}C{2cm}L{9cm}}
			\cmidrule{1-4}    \multicolumn{1}{c}{\textbf{Profile}} & \textbf{Sample size} & \textbf{Interview numbers} & \multicolumn{1}{c}{\textbf{Background}} \\
			\midrule
			Contributor IEC & 1 & 13    & \multicolumn{1}{l}{IACS software life cycle standardisation} \\
			\midrule
			Contributor SAFe & 1 & 12    & \multicolumn{1}{l}{IACS agile development} \\
			\midrule
			Principal Expert & 3 & 4, 5, 8 & IACS security standards and processes, security life-cycle, security architecture \\
			\midrule
			Senior Expert & 4 & 1, 2, 6, 9 & Cloud security, methods and tools for secure solutions, cyber security coaching, security processes improvement, IT security assessments \\
			\midrule
			Expert & 7 & 3, 7, 10, 11, 14, 15, 16 & IACS agile development, quality compliance, development of access control systems, data privacy on smart cities, security design management, DevOps, security tools development, automated security testing, IT security in critical infrastructure \\
			\bottomrule
	\end{tabular}}%
	\label{tab:6_evaluatorsPerProfileGroup}%
\end{table}%

\subsection{Survey Instrument}

Since our goal is to explore practitioners' opinions about \ssafe, we identified semi-structured interviews as the most suitable technique~\cite{Shull:2007}. Each interview lasted 1.5 to 2 hours and took place in an isolated environment with one interviewee and two interviewers. One interviewer actively followed the questionnaire and the other one documented the answers and controlled attachment to interview protocol, available at our online material.

\begin{figure}[ht]
	\centering
	\includegraphics[width=0.8\textwidth]{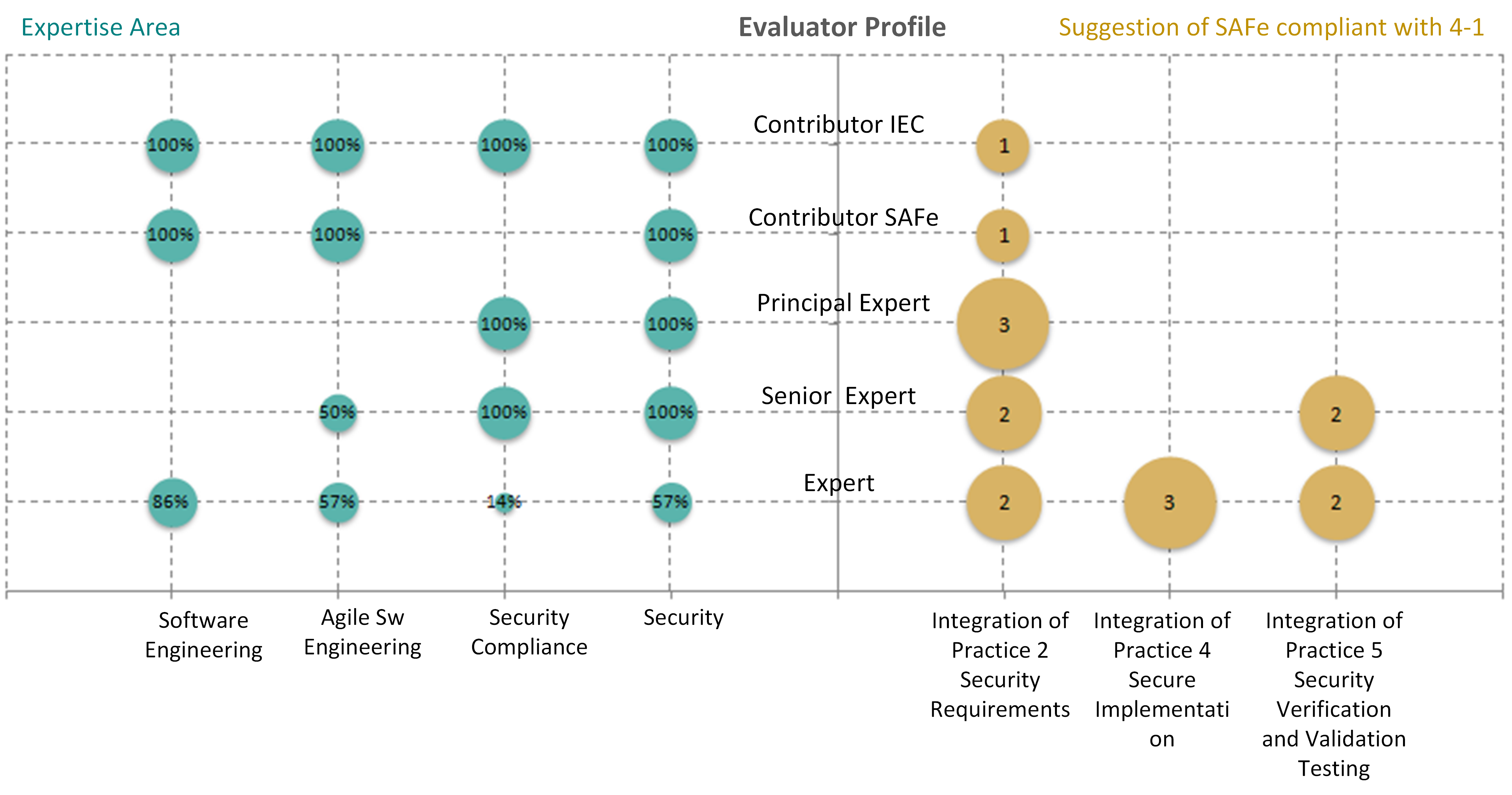}
	\tiny \caption[\ssafe suggestions distribution into profile groups]{\small \ssafe Suggestions distribution into profile groups. Right side: number of interviewees per suggestion. Left side: percentage of interviewees per expertise area.}
	\label{fig:6_expertiseAndSolution}
\end{figure}

Each interview was dedicated to one \ssafe element according to the subject's background: SR, SI, or SVV (c.f. \cref{sec:method}). Subjects were also introduced to the \ssafe CDP to have an overview of the processes involved the framework as shown in \cref{fig:6_expertiseAndSolution}. 

Interviewers first briefed individual subjects about the interview flow and the purpose of \ssafe models as well as their hierarchy (overview model and individual practice models) but did not provide any instruction or training on the actual models. 
Then they showed a textual excerpt from 4-1 and SAFe, followed by the corresponding individual models and finally merged models from \ssafe. Subjects rated the perceived usefulness and practical applicability of each representation. Notes from throughout the interview were discussed before the interview's end to complete the picture.

%\subsection{Data Analysis}

%Evaluation is based on summarising the answers to closed questions and clustering comments and concerns according to commonalities. We further analysed the emphasis of answers to differentiate acceptance vs. conviction, rejection vs. repulsion, and neutrality vs. doubt. Hence, we tabulated answers according a 9-point Likert scale.

\section{Study Results}
Evaluation is based on summarising the answers to closed questions and clustering comments and concerns according to commonalities. We further analysed the emphasis of answers to differentiate acceptance vs. conviction, rejection vs. repulsion, and neutrality vs. doubt. Hence, we tabulated answers according to a 9-point Likert scale.
In the following, we summarise and interpret our results according to our research questions.

\subsection{Subject Knowledge}

\begin{figure}[!htbp]
	\centering
	\includegraphics[width=0.7\textwidth]{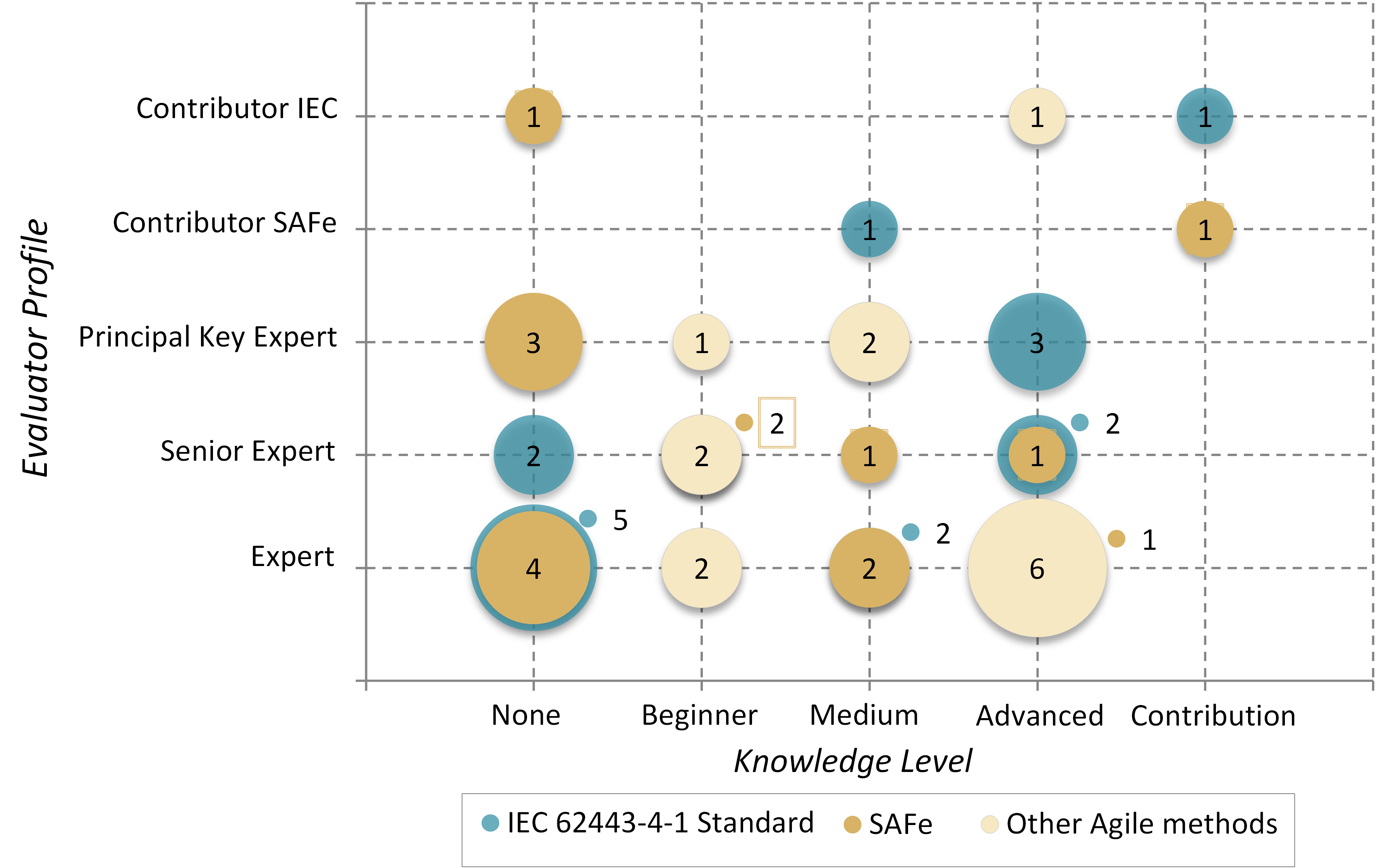}
	\small \caption[Subject knowledge of SAFe and 4-1]{Subject knowledge of IEC~62443-4-1 and SAFe or comparable process frameworks.}
	\label{fig:6_SAFeAnd41}
\end{figure}

In total we selected 16 subjects with different levels of knowledge about 4-1 and SAFe. \Cref{fig:6_SAFeAnd41} shows that now all of them know 4-1 but all except one are aware of other security and safety standards such as ISO/IEC~27001 or other standards of the IEC~62443 family. Similarly, not all know SAFe but all are familiar with other agile process frameworks such as Scrum.

\subsection{Applicability of \ssafe (RQ~1)}{Applicability of S2C-SAFe (RQ~1)}
\label{sec:evaSolution}

We consider two aspects: applicability itself and potential implementation problems.
Overall, while all interviewees strongly agree on the potential of using the integrated model as a means to foster discussions with their counterparts, they see potential problems in the integration of security aspects.

\subsubsection{Applicability} \ \\

\begin{figure}[!htbp]
	\centering
	\includegraphics[width=0.8\textwidth]{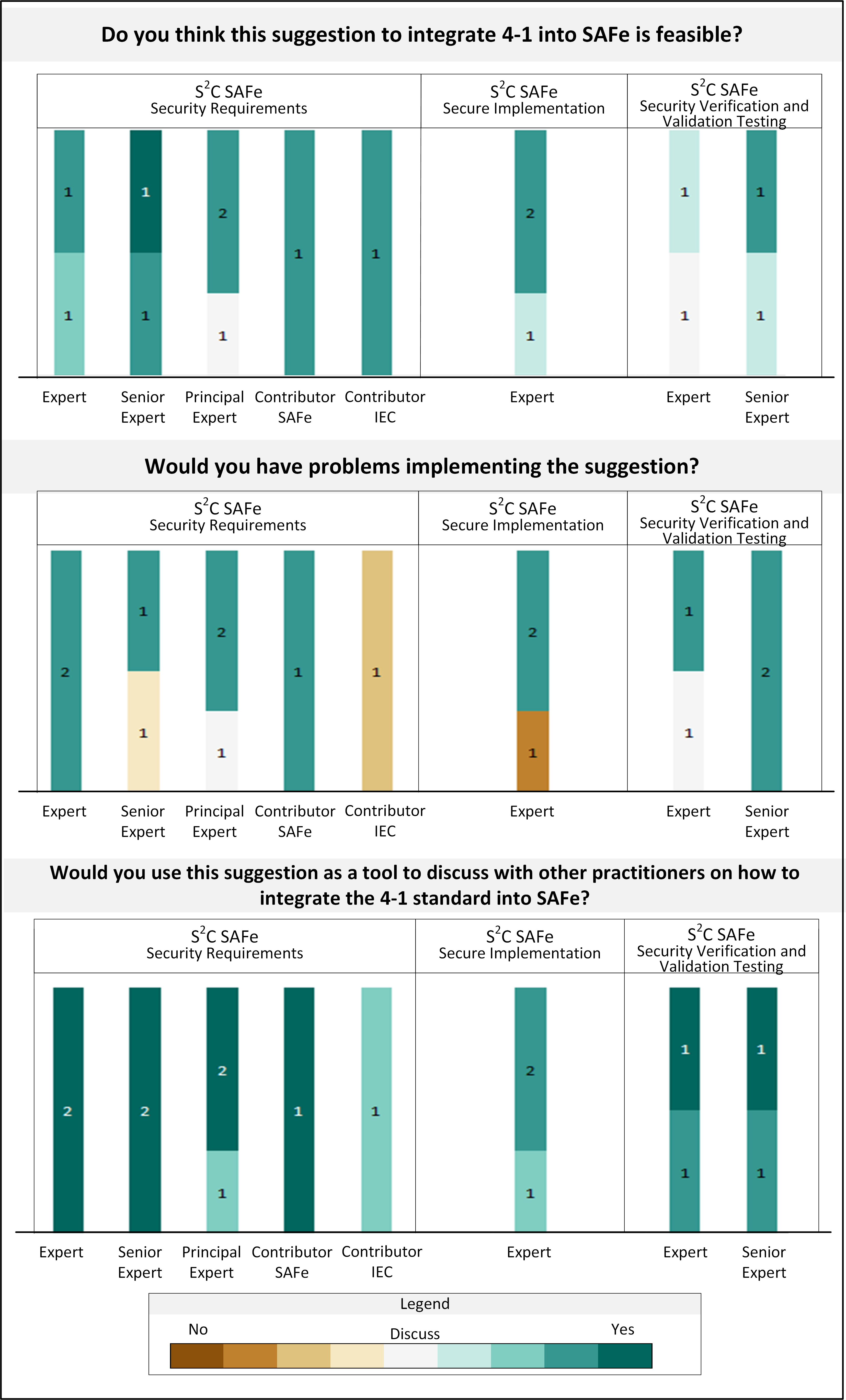}
	\small \caption[Summary of opinions on \ssafe applicability]{Summary of opinions about \ssafe applicability based on suggestions regarding 4-1 practices.}
	\label{fig:6_Answer4}
\end{figure}

\ssafe demonstrates that SAFe can be compliant with the 4-1 standard. All interviewees deem it usable in their environments and expressed their desire to use it for discussion with other practitioners (see \cref{fig:6_Answer4}). They particularly stated that it would provide a common language between security and development fields; some even saw it as the only such tool they are aware of (see \cref{tab:6_clusters4}). The following paragraphs give detailed results for each of the 4-1 practices introduced in \cref{sec:method}.

\textbf{SR:} 
Subjects strongly agree that this suggestion is feasible. A \emph{principal expert} (\#8) did not give a positive answer, but instead argued about the complexity of having security experts within teams in general. Almost all envision problems during implementation, most relate to the lack of security practitioners, team security awareness, or split security requirements. \emph{Contributor SAFe} thinks that proposed security activities overload PI planning while \emph{contributor IEC} sees no problems if models are shown only to people that design processes and not to agile team. However, all subjects plan to use the suggestions as a discussion tool with their respective counterparts.

\textbf{SI:} 
Subjects strongly agree that this suggestion is feasible. One DevOps \emph{expert} (\#7) argues that educating the product owner on security is complex. Instead they propose a \quotes{security product owner} who would be capable of extending the definition of done (DoD) with security aspects. In contrast, an \emph{expert} product owner (\#14) remarks the adapted DoD as a key to apply. An \emph{expert} security consultant (\#11) is confident that problems would exist although they cannot refer to any specific one.

\textbf{SVV:}
Although overall positive, opinions on feasibility of this suggestion are not as decided as previously. Two respondents (\#11 \emph{expert} scrum master and \#1 solutions security \textit{senior expert}) find the suggestion feasible and well integrated. Another security \textit{senior expert} (\#6) is concerned about automation support for testing non-functional security aspects and about effort for security practitioners. A security assurance \emph{expert} (\#3) argues about the role and interactions of security practitioners throughout the process. Hence, all of them envision problems related to the integration of automatic testing, workload, and expertise of security practitioners.

Additionally, as interviewers we experienced that \ssafe improves communication among practitioners with different profiles and backgrounds. We actively discussed interviewees' issues on security and agile development. All explanations were based on the models we provided. Subjects with the highest level of knowledge (\emph{Contributor IEC} and \emph{Contributor SAFe}) challenged us with management or operational questions, e.g., how to implement or even potentially bypass certain aspects. We succeeded in explaining our perspective purely by pointing out specific model aspects. Conversations were dynamic, indicating a common understanding between interviewer and interviewee.
\Cref{tab:6_clusters4} summarises key opinions on \ssafe while \cref{tab:6_quotesBenefit} lists noteworthy remarks.

\begin{table}[!htbp]
	\centering
	\caption[Evaluation of \ssafe: common issues]{Summary of key opinions on \ssafe.}
	\scalebox{0.9}{	\begin{tabular}{L{14cm}L{2.3cm}}
			\toprule
			\multicolumn{1}{l}{\textbf{Opinion}} & \multicolumn{1}{l}{\textbf{Interviews}} \\
			\midrule
			Facilitates common language to discuss between security experts and agile team  & 2, 5, 14, 16 \\
			\midrule
			Solution is a comprehensible, clear guide & 4, 5, 7, 8 \\
			\midrule
			Increment effort and workload & 5, 6, 9, 12  \\
			\midrule
			Concern about roles expertise to accomplish tasks: Product Owner, Product Management & 3, 4, 6, 10  \\
			\midrule
			Need to increase security awareness & 1, 3, 7, 8  \\
			\midrule
			Concerns on expertise and profile of security experts & 1, 4, 10 \\
			\midrule
			To have security practitioners within agile teams is challenging& 8, 10  \\
			\midrule
			Need to have a deep understanding of own process to implement suggestion & 1, 16  \\
			\midrule
			It is the only tool available & 7, 11  \\
			\midrule
			Concerns about fit activities into short cycles & 8, 12  \\
			\midrule
			Find color-coding is useful & 7, 13 \\
			\bottomrule			
	\end{tabular}}
	\label{tab:6_clusters4}%
\end{table}

% Table generated by Excel2LaTeX from sheet 'Quotes'
\begin{table}[!htbp]
	\centering
	\caption{Interviewees' statements on \ssafe.}
	\scalebox{0.9}{
		\begin{tabular}{L{8.5cm}C{1cm}L{2.5cm}L{3cm}}
			\hline
			\textbf{Quote} & \textbf{Interviews} & \multicolumn{1}{c}{\textbf{Profile}} & \multicolumn{1}{c}{\textbf{Background}} \\
			\hline
			Big advantage, we could speak same language as SAFe experts. This would dramatically reduce problems to adapt SAFe.  Yes, I would love to use it as a discussion tool. & 2 & {Senior Expert} & {Security compliance} \\
			\hline
			It makes sense what you did. If it is not possible SAFe is broke & 4 & {Principal Expert} & {Security research} \\
			\hline
			Sure is feasible, how to measure success I wonder & 5 & {Principal Expert} & {Head security group} \\
			\hline
			It is a very nice way to reduce complexity to discuss. & 7 & {Expert} & {DevOps } \\
			\hline
			Visibility of security into agile development environment. Transparency of what is being achieved & 9 & {Senior Expert} & {Security assessments} \\
			\hline
			Sure, there is nothing else. I don't think there is anything & 11 & {Expert} & {Security consultant / Scrum practitioner} \\
			\hline
			We need to involve a pilot implementation & 12 & {Contributor SAFe} & {Head development group} \\
			\hline
			I will use it as a basis to communication
			& 14 & {Expert} & {Product Owner} \\
			\hline
			I like it. It makes dedicated to think about security & 15 & {Expert} & {Systems Architect} \\
			\hline
	\end{tabular}}
	\label{tab:6_quotesBenefit}%
\end{table}%
\vspace{-20pt}
\subsubsection{Potential Implementation Problems} \ \\
Our interviewees raised concerns regarding implementation of \ssafe in their project settings. They are particularly interesting to us as they help steering future adaptations and because some concerns are rather general challenges on the integration of security, let alone continuous security engineering. These concerns can be summarised as follows:

\paragraph{Models should guide instead of comprehending compulsory processes} \ \\
One \textit{senior expert} argued that if a model is too strict, people will not adapt it and bypass compliance efforts (\#1). The suggestions seem difficult to implement in iterations or in specific program increments. This seems particularly true for security testing (vulnerability/penetration) prior to or during a \textit{System Demo}. This highlights the need for an incremental prototypical implementation of individual suggestions to shed light on potential adaptation barriers which might differ in dependency to the practices and the roles. 

\paragraph{Achievement of security expertise and awareness}\ \\
During the design phase, we emphasised that \ssafe cannot compensate for a Product Owner with knowledge of 4-1. Our interviewees confirm that this holds not only for general SAFe roles but also for security practitioners in general. Both security and agile development experts agree that security expertise for each part of the solution requires specialisation. Such specificity on profiles would aggravate the deficit of security professionals.  Exemplary statements are \quotes{During verification of compliance, people tend to deviate from the standard} (\#7) or \quotes{Lack of experience on security compliance leads to failed projects} (\#3, security \emph{expert}).

\paragraph{Difference between agile and express development delivery}\ \\
Security is generally perceived to be something that slows down agile development processes. Some exploratory questions revealed that agile time constraints are not followed in our settings, e.g., daily meetings last more that 15 minutes. Our concept of agile therefore seems to relate more to iterative and incremental development than to express delivery and integrating security-related activities will surely expand this gap further. While we understand the need for a trade-off between effort and cost for adapting security (or any other quality facet) this aspect seems particularly hard to achieve and constitutes an open issue.

\subsection{Continuous Security Compliance Challenges  (RQ~2)} \label{sec:challengesSecurity}

The interviewees were asked to mention priorities among the security activities described in the 4-1 development life cycle. \textit{Security requirements} (SR) seems to be the most challenging practice for our interviewees. Other priorities differ per profile, as shown in the examples for \textit{security management} and \textit{security verification and validation testing}.   

The top priority issue is raising awareness for security to achieve continuous security compliance. Second place is taken by an adequate prioritisation of security aspects and common perspectives among management and teams. Challenges for security integration into continuous software engineering seem similar to those with linear development models. Subsequently we summarise our key findings on the challenges raised.

\paragraph{Security requirements elicitation:} Challenges go beyond elicitation, from prioritisation over allocating them to increments and tracking adequate testing. Respondents extend the concern to overall 4-1 activities into cycles, e.g., threat analysis, testing, or issue management. Some related quotes include \quotes{What does the standard says about iterations and when the required process should occur again?}(\#15, software architect) or \quotes{Problem is to identify what is the most important and which things can be done in parallel} (\#12).

\paragraph{Security more than a non-functional requirement:} 4-1 contains an overview of security as described by compliance. Our interviewees state that security is normally addressed via functional requirements while other aspects, such as management-related ones, are too often left behind.

\paragraph{Software architecture impact:} Software architectures are built incrementally in continuous development. One interviewee argued in particular: \quotes{How to have security design or requirements of something we don't know yet, something we create on the go} (\#12, \textit{Contributor SAFe}). We argue that security analysis can be done while thinking about the goal and later iteratively extend it to the solution-specific components. However, this needs a certain continuity just like other non-functional properties, which project participants seem to see as difficult to achieve.

\paragraph{Improvement demand for security expertise and awareness:} In development teams the lack of expertise for security seems to be a common theme~\cite{Bartsch:2011}. Particularly, our group of interviewees seems to have a sound level or security awareness: \quotes{I see the need of security} (\#15, product owner). They comprehend that challenges also depend on the role and therefore some interviewees even suggest to define new (agile) security-related profiles such as a \quotes{secure product owner} or a \quotes{secure system architect}. Furthermore, respondents argue that security expertise should generally be improved to achieve compliance. This is exemplified in the following quote: \quotes{A new secure product owner could do it} (\#7). %\quotes{\textit{Lack of experience of security experts lead to failed projects}} (I3-security \emph{expert}).
Interestingly, these observations corroborate the need to raise a common awareness for security in the overall agile team: \quotes{implementations deviate from standard [and often] lead to fake implementations} (\#2, security compliance \textit{senior expert}); \quotes{There are guidelines to bypass compliance rules} (\#8, security \textit{principal expert}).

\paragraph{Security compliance as a common agreement:} Related to our previous observation is that subjects perceive compliance as a complex endeavour. They noticed that management, teams, and even compliance practitioners have different perspectives on compliance. Some see security compliance as a burnout journey, others as a luxury and others again as a worthwhile goal. A common agreement on the need to achieve common security standards is therefore a prerequisite for the success of our undertaking.

\paragraph{Misunderstandings of agile engineering terms:} In our interviews we noticed that terms are used often in a cumbersome manner. For instance, subjects with agile development knowledge (e.g., \#1, \#2, \#3) often referred to Scrum only implicitly by mentioning specific elements such sprint, iteration, and product owner; \quotes{definition of done} was often used when referring to acceptance criteria; other interviewees had difficulties in capturing the notion of artefacts in context of process models: \quotes{the word artefact is not easy} (\#10, \emph{expert}). As a matter of fact, such key concepts are still subject to current debates and need further attention in future work generally dealing with software processes~\cite{Mendez:2018}.

\section{Conclusion} \label{sec:conclusion}
%\subsection{Summary of conclusions and Key Takeaways}

%Although agile development is the most adopted method, it is not well integrated with security yet~\cite{Turpe:2017,ahola_handbook_2014,Fitzgerald:2017,Bell:2017}. Other approaches involve security based on particular security-related concepts like vulnerability testing~\cite{Stephanow:2017}, coding analysis \cite{Baca:2015}. Our contribution approach security through compliance --- implementing security considering the triad: people, process and technology. This work fills a research gap of a step by step solution to integrate a security standard into a large scaled agile method.
In this paper we reported on our work towards integration of security requirements derived from IEC~62443-4-1 into large-scale agile development based on SAFe in order to facilitate CSC. We presented the \ssafe framework and evaluated it based on interviews with 16 industry experts. Evaluation results strengthen confidence that this approach and the resulting models provide a feasibly way for security compliance in large-scale organisations practising lean and agile development.

\subsection{Impact and Implications}

Results show \ssafe models have a clear impact for practitioners. They show precisely how software engineering and security practitioners have to interact to achieve the goal of security compliance. Furthermore, the models can be understood in a time-effective manner and challenge popular belief that agile processes are a gateway to chaos and therefore not reconcilable with security and compliance concerns.
The unanimous response to our work was the exact opposite: Introducing large-scale agile processes demands a culture and mindset change. Even though not our intention, the models helped to convey to sceptical practitioners that both secure and agile development is feasible at scale with reasonable effort. 

Our research strongly indicates that models are an excellent way to mediate between agile practitioners and security experts. Particularly visual models allowed them to engage the challenge of continuous security compliance together. Moreover, these models pave the way for analysing various further challenges of the research field: Do models increase the speed of adapting large organisations to secure agile processes at scale? Are models a better way of getting security norms accepted in daily software engineering activities? Can models provide guided and precise support for secure agile security governance? We are confident that our contribution supports researchers to further investigate these questions. 
%Important is we pursue not only secure agile development (continuous security) but security compliant agile software engineering (continuous security compliance). Other solutions target agile security which means software assurance but no regulation compliance. This solution integrates standard requirements without burning out agile methods while solving security concerns from practitioners and researchers. Therefore both assurance and compliance certification are achieved.

\subsection{Relation to Existing Evidence}

Our study is in tune with existing trends of empirical studies on secure software engineering~\cite{Othmane:2017}, but extends the study population in number and profile. To the best of our knowledge, preceding studies involved up to 11 practitioners with mixed background or students as subjects and focused on valuated, yet isolated topics. An integrated view on a security standard compliant agile framework was not in their scope. Our contribution is aimed at this gap and involves 16 experienced professionals, partially with contributing roles to the standards or decision-making roles in the organisation. We focused on the highest ranking experts available. As explained, a SAFe integration may last up to 8 years and the interviewees are high-ranked professionals. Their opinion is the closest to certainty in a timely evaluation. 

\subsection{Limitations and Threats to Validity}
\label{sec:evaLimitations}

Qualitative studies inherently carry limitations and interview research in particular has threats to validity that need discussion, the most important of which shall be discussed here.

The individual expertise of each participant might influence their attention and interpretation of security requirements as well as agile practices captured in the models. We tried to mitigate this with discussion-intensive preparation procedures, but also by letting subjects interpret the models as they are without any further instruction. We were interested in potential bias towards the subject of security compliance as that reflects on the projects where those models shall be applied.

Similarly, involving experts from each respective field carries the risk of self-selection and confirmation bias. To mitigate this we selected subjects according to typical roles in the target organisation environment instead of their particular interest in the topic. The same is true for which part of \ssafe they reviewed (requirements, implementation, or testing). We also designed interview plan and questionnaire accordingly and allocated interviewees to models based on previously defined profiles.

%\paragraph{Model extension} \ssafe incorporates artefacts in context of a larger process. It is necessary to refine them to the a level that allows to incorporate templates for direct application at project level. This is particularly important to increase the awareness of what to document (artefact templates), roles and responsibilities (profiles), and finally overcoming potential barriers resulting from the lack of security expertise (guidelines). A further aspect worth considering is the possibility to assess the process outcomes in order to obtain security certifications based on the documented results.

%\paragraph{Scaling up to practice} So far, we have conducted qualitative studies to test the sensitivity of our model to practical contexts. The insights we received help us already steering future refinements of our model. The next step consists of applying our refined model in context of a longitudinal study that allows us to better understand the larger implications of a new model in context of large-scaled agile projects. 

Overall, our study already strengthens our confidence in the capability of \ssafe to integrate security and compliance concerns with lean and agile development. We cordially invite researchers and practitioners to join our endeavour towards facilitating continuous security compliance in large organisations and regulated environments.

\section{Acknowledgements} To the practitioners that evaluate this work, as well as to Markus Voggenreiter and Florian Angermeir for reviewing.
\newline 

\begingroup
\let\clearpage\relax
\bibliographystyle{splncs04}
\bibliography{bibliography.bib}
 \endgroup
  \clearpage
\end{document}